\documentclass[]{raa}
\usepackage{deluxetable}
\usepackage{amssymb}            % referee version: for submission
\usepackage{graphicx,times}
\usepackage{natbib}
\usepackage{color}

\providecommand{\bm}[1]{\mbox{\bol
dmath$#1$\unboldmath}}

\begin{document}

   \title{Statistical studies of optically dark gamma-ray bursts in the $Swift$ era}

 \volnopage{ {\bf 2009} Vol.\ {\bf 9} No. {\bf 10}, 1103--1118}
   \setcounter{page}{1}

   \author{Wei-Kang Zheng
      \inst{1,2}
   \and Jin-Song Deng
      \inst{1}
   \and Jing Wang
      \inst{1}
   }

   \institute{National Astronomical Observatories, Chinese Academy of Sciences,
             Beijing 100012, China; {jsdeng@bao.ac.cn}\\
        \and
             Graduate University of Chinese Academy of Science, Beijing 100049,
             China \\
%        \and
%             Full institute address for the third author
\vs \no
   {\small Received 2009 April 29; accepted 2009 June 6}
}

\abstract{ We compare the properties of optically dark GRBs, defined
by the optical-to-X-ray spectral index $\beta_{\rm OX}<0.5$, and normal
ones discovered by the $Swift$ satellite before the year 2008 in a
statistical way, using data collected from the literature and
online databases. Our sample include 200 long bursts, 19 short bursts,
and 10 with measured high redshifts ($z\gtrsim 4$). The ratio of dark
bursts is found to be $\sim 10-20\%$, and is similar between long bursts,
short ones, and the high-$z$ sub-sample. The result for long bursts
is consistent with both the pre-$Swift$ sample and studies by other
authors on smaller $Swift$ samples. The existence of dark short GRBs
is pointed out for the first time. The X-ray derived hydrogen column
densities of dark GRBs clearly prefer large values compared with
those of normal bursts. This supports the dust extinction scenario
as the main cause of dark GRBs. Other possibilities like very high
redshifts and non-standard emission mechanisms are less likely although
not fully excluded.
 }

   \authorrunning{W. Zheng et al. }            %author_head in even pages
   \titlerunning{Statistical Studies of Optically Dark GRBs in the $Swift$ Era }  % title_head in odd pages
   \maketitle

%% The author head (on even pages) and the title head (on odd pages) will be
%% automatically extracted from \author{} and \title{}. Whenever the title is too long,
%% you will be asked to supply a shorter one by inserting either \authorrunning{} or
%% \titlerunning{} before \maketitle. Anyway, you can specify your own heads.
%%
%%
%% Note: In the following text body of your manuscript, please note several differences from
%%       other major journals:
%% (1) \subsection{Please Capitalize the First Letter of Each Notional Word in Subsection Title}
%% (2) Please Capitalize the First Letter of Each Notional Word in all tables' captions

%
%________________________________________________ sections below
%
\section{Introduction}           %% first-level sections will be auto-capitalized
\label{sect:intro}

There has been a long-standing problem of optically ``dark'' Gamma-Ray
Bursts (GRBs) since the discovery of afterglows. Nearly 90\% of
well-localized GRBs have an identified X-ray afterglow (De Pasquale
et al. 2003; Gehrels et al. 2007). However, some of them elude an
optical detection. In the $BeppoSAX$ satellite era, over $\sim
60-70\%$ of GRBs with optical follow-ups failed to show an optical
counterpart (Fynbo et al. 2001; Lazzati et al. 2002). The long time lag
between BeppoSAX detection and the optical follow-up was suspected to be
at least partly responsible since a GRB afterglow decays rapidly in
brightness. The HETE-2 mission, with better GRB localizations and
rapid coordinate dissemination ability, reduced the optical
non-detection fraction to $\sim$ 10\%, once claimed as the end of
the ``dark burst'' mystery (Lamb et al. 2004).

The problem resurfaced quite unexpectedly in the $Swift$ era. The
Burst Alert Telescope of the satellite can localize a triggered GRB
with high precision ($\sim$ 3') almost instantly (Gehrels et al.
2004). The X-Ray Telescope and Ultra-Violet/Optical Telescope
onboard routinely perform follow-up observations starting from just
a few minutes after the onset of a GRB. A large global network of ground
fast-response telescopes, on alert for a BAT GRB trigger, can slew
for optical follow-ups within minutes (e.g., Zheng et al. 2008),
followed then by big telescopes used for deep searches. Precise XRT
localization ($\sim$ 5") helps in optical counterpart
identifications. However, defying all those advantages, the
detection rate in early UVOT observations ($<1$ hr) turned out to be
only $\sim 30\%$ (Roming et al. 2006). Recently, Melandri et al.
(2008) and Cenko et al. (2009) found a significant incidence of dark
bursts in their respective systematic ground follow-ups, which went
deeper than the UVOT. Those authors utilized a more physical
definition of dark bursts (Jakobsson et al. 2004), which evaluates
the optical brightness relative to the X-rays.

Several suspects have been proposed as the cause of optical
darkness. 1) For a GRB at very high redshift ($z\gtrsim 4$),
the hydrogen Ly$\alpha$ blanketing and absorption of intervening
host-galaxy or intergalactic medium may greatly suppress the flux in
an observer-frame wavelength below $1215(1+z)$~\AA\, (Lamb \&
Reichart 2000). However, GRBs with such a confirmed high redshift are
still rare. 2) Extinction by dust in the host galaxy of a
dark burst may be very high (e.g., Djorgovski et al. 2001; Rol et
al. 2007; Jaunsen et al. 2008), working together with the relatively
high GRB redshift which may shift UV photons, which have much
larger extinction coefficients than optical ones, into the observing
optical band (Klose et al. 2003). 3) The optical light curve may
have decayed very rapid before the optical search started (e.g.,
Berger et al. 2002), a scenario seemingly no longer relevant in the
$Swift$ era. 4) Last but not least, some GRBs may have an
intrinsically low efficiency in optical afterglow emissions (e.g.,
Jakobsson et al. 2005; Oates et al. 2006; Urata et al. 2007)
relative to X-rays, lower than that predicted by the standard afterglow
model.

In this paper, we study the large sample of GRBs detected by the
$Swift$ satellite up to the end of the year 2007, comparing the
statistical properties between optically dark bursts and normal
ones. Our sample and data selection are described in Section
\ref{sect:sample}, and detailed analysis is in Section \ref{sect:DataAnalysis}.
Implications of our results for the various dark burst scenarios
are discussed in Section \ref{sect:Discussions}. We compare our
studies with similar ones in Section \ref{sect:Comparison}.

% Authors can give a citation as `Michel et al. 1992'.
% You may also use \cite, \citep and \citet for citation, and use Table~1
% or Figure~1 and so forth. Using \ref and \label for cross-references of
% Tables/Figures is a good way in adjusting/adding/removing text, tables or
% figures.

\section{The GRB Sample and Data Selection}
\label{sect:sample}

We selected a sample from the GRBs detected by $Swift$ BAT up
to the end of the year 2007, collecting the redshift, BAT fluences between
$15-150$~keV, $R$-band flux densities and X-ray integral fluxes of
$0.2-10$~keV at 11~hr after the BAT trigger, and intrinsic hydrogen
column density $N_{\rm H}$. The $0.2-10$~keV X-ray integral flux was
converted to the flux density at 3~keV using the measured X-ray
spectral index (or using the mean value $1.1$ if no measured value
was found). Optical fluxes have been corrected for Galactic
extinction (Schlegel et al. 1998). We derived the optical-to-X-ray
spectral index at 11~hr, $\beta_{\rm OX}$
($F_{\nu}\propto\nu^{-\beta}$), from the ratio of  the $R$-band flux
density to the 3~keV one, in order to distinguish the optically dark
bursts from the GRB sample (Jakobsson et al. 2004).

For the sake of convenience, we labeled our GRBs with optical
detections as ``OT'' and those with only optical upper limits as
``UL''. We excluded GRBs without a detected X-ray afterglow
since a measured X-ray flux is essential for our dark burst
definition.

We chose Nysewander et al. (2009) as our main data source except for
the $N_{\rm H}$ values. Those authors made a comprehensive data
compilation of the GRBs with X-ray or optical follow-up observations
detected before the year 2008 by various satellites and having, after
a thorough literature search to find the best data. Note that
if no 11-hr $R$-band or X-ray observational data were available,
data extrapolation was done by them. For that purpose, a power-law
temporal decay of the X-ray flux, with a slope of $1.2$, that of the
optical flux, with a slope of $0.85$ for long bursts and of $0.68$
for short bursts, and an optical spectral index of $1.0$ were
assumed. A similar compilation made by Gehrels et al. (2008) for the
$Swift$ GRBs only extends to 2007 July. For some of the GRBs observed
by Melandri et al. (2008), extrapolating the $R$-band upper limits
reported there (usually of combined images) to 11~hr results in values
significantly deeper than in Nysewander et al. (2009), so those
deeper values were adopted. We replaced the $R$-band slope value
$1.1$, which was assumed in Melandri et al.(2008), with $0.85$ when
doing extrapolation in order to maintain data uniformity. Finally, we
found three $Swift$ bursts, GRB 050502B, GRB 060602B, and GRB
060807, that were somehow missed from Nysewander et al. (2009). We
included two of them, collecting data from other literature, but we
rejected GRB 060602B which suffered from incredibly large Galactic
extinction.

Two kinds of intrinsic $N_{\rm H}$ values were compiled by us. Both
were derived from X-ray spectral data-reductions taking into account
the host-galaxy gas absorption, with solar metallicity assumed, while the
Galactic $N_{\rm H}$ was fixed to the value of Dicky \& Lockman
(1990). For one kind, i.e., $N^0_{\rm H}$, the redshift of the host
galaxy was simply fixed at zero. For the other, the measured
host-galaxy redshift was adopted and we designated the fitting
intrinsic column density as $N^z_{\rm H}$. We first collected the
$N^z_{\rm H}$ values in Grupe et al. (2007), followed by more from
the UK Swift/XRT GRB spectrum repository
\footnotemark{}\footnotetext{http://www.swift.ac.uk/xrt\_spectra/}
(Evans et al. 2007), and the rest from other literature. For the
GRBs without measured redshifts, the UK repository provided us with
the $N^0_{\rm H}$ values, while for those of known $z$, we searched the
GRB Coordinates Network \footnotemark{}\footnotetext
{http://gcn.gsfc.nasa.gov/} (GCN) Circulars and Reports for
$N^0_{\rm H}$. The UK repository had not yet responded to a few
cases of late reported $z$ by the time this paper was written. For
those, we multiplied $N^0_{\rm H}$ by $(1+z)^{2.1}$ to get empirical
$N^z_{\rm H}$ values. The fitting $N_{\rm H}$ for the XRT Windowed
Timing data and that for the Photon Counting data could be different
if both were available; we chose the latter, as Butler \& Kocevski
(2007) suggested that it may better reflect the true value. In some
cases, no gas absorption above the Galactic value was needed at all,
so the intrinsic $N_{\rm H}$ value had to be left open.

Our final sample, which consists of 229 $Swift$ BAT GRBs in total,
is listed in Tables 1 and 2. Among those, we seek out the ten with
measured high redshifts ($z\gtrsim 4$) to make a high-$z$ sub-sample
(Table \ref{tab:HighZEvents}), whose $R$-band fluxes must have been
significantly diluted by intergalactic H Lyman blanketing absorptions.
The remaining 219 include 200 long bursts ($T_{90}>2$~sec; Table
\ref{tab:SwiftBursts}) and 19 short bursts ($T_{90}<2$~sec; Table
\ref{tab:SwiftBursts}). Note that all the high-$z$ sub-sample members
are also long bursts.

Following Jakobsson et al. (2004), we define dark bursts as those with
$\beta_{\rm OX}<0.5$. This definition has operational advantages
over the ``optical non-detection'' one that was typically described
by a more or less arbitrary threshold in the limiting apparent
magnitude (e.g., $R\sim 23$ within 2 days; Djorgovski et al. 2001).
More importantly, it has a clear physical meaning. The standard
afterglow model assumes that there is synchrotron radiation from
relativistic electrons accelerated by the forward-external shock. For slow
cooling which is certainly the case at an epoch as late as 11 hr,
the X-ray or optical spectrum above $\nu_c$ must be
$\propto\nu^{-p/2}$, and that of $\nu_m<\nu<\nu_c$ proportional to
$\propto\nu^{-(p-1)/2}$ (see Zhang \& M\'{e}sz\'{a}ros 2004 for a
review). Below $\nu_m$, the spectrum can rise to higher frequencies
but a corresponding rising light curve has not been observed at a late
time, so this case can be excluded. The electron spectral index $p$
is typically $\sim 2-2.5$. As for $p>2$, which is most reasonable
in shock acceleration physics, $\beta_{\rm OX}$ cannot be smaller
than 0.5 in order to be compatible with the standard afterglow
model. To allow for some measurement errors, we designated the
boundary cases, i.e. with $0.5<\beta_{\rm OX}<0.6$, as ``gray''
bursts. Rol et al. (2005) proposed another method to identify dark
bursts by comparing optical fluxes/upper limits with model
extrapolations from the X-rays for any specific GRB. This may be more
precise than using $\beta_{\rm OX}$, but it seems difficult to
implement for a large sample.

\section{Data Analysis}
\label{sect:DataAnalysis}
\subsection{The optical flux versus X-ray flux diagram}\label{sect:OptXray}

\begin{figure}[]
\centering
   \includegraphics[bb=45 160 565 580,width=0.7\textwidth]{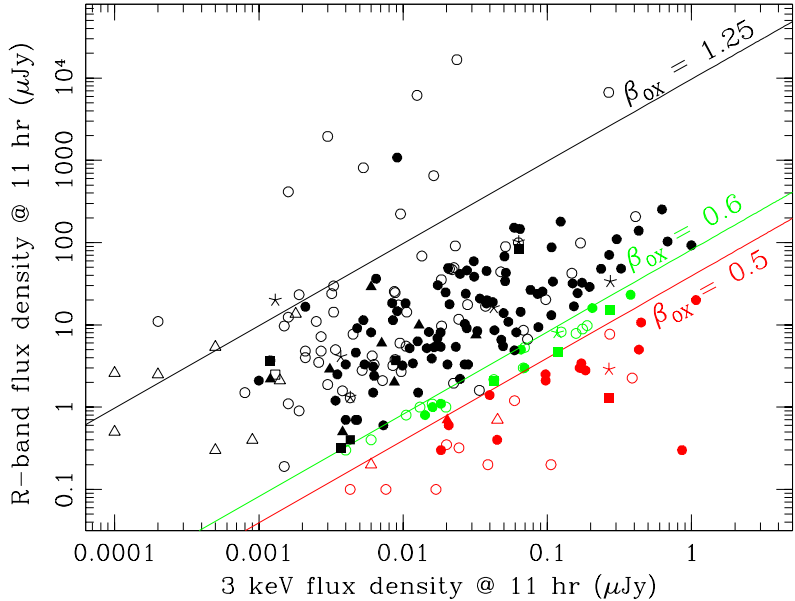}
   \caption{$R$-band flux density versus the 3~keV X-ray flux density
at 11~hr. Dark bursts ($\beta_{\rm OX}<0.5$) are drawn in {\it red},
those of $0.5<\beta_{\rm OX}<0.6$ in {\it green}, and the others in
{\it black}. {\it Triangles} present short bursts, and {\it circles}
long bursts. A {\it filled} symbol means optical detection being
reported (labeled OT), while an {\it open} one indicates optical upper limits
only (labeled UL). For the 10 high-redshift bursts ($z\gtrsim 4$),
two values are given for each (see \S \ref{sect:HighZ}), one {\it
square} and the other {\it asterisk}. Also plotted are lines
indicating $\beta_{\rm OX}=0.5$ ({\it red}), 0.6 ({\it green}), and
1.25 ({\it black}), respectively.\label{fig:XversusOptFlux}}
\end{figure}

The $R$-band flux densities and the 3-keV X-ray flux densities of
our sample are plotted in Fig. \ref{fig:XversusOptFlux}. Both long
bursts ({\it circles}) and short bursts ({\it triangles}) are
included. For our high-$z$ sub-sample ($z\gtrsim 4$), two kinds of
$\beta_{\rm OX}$ values are shown ({\it squares} and {\it
asterisks}; see \S \ref{sect:HighZ}). OT events are denoted by {\it
filled} symbols, while UL events by {\it open} ones. Three lines are
drawn to show the slope positions corresponding to $\beta_{\rm
OX}=0.5$ ({\it red}), 0.6 ({\it green}), and 1.25 ({\it black}),
respectively. Most of our sample is confined between $\beta_{\rm
OX}=0.5$ and $\beta_{\rm OX}=1.25$, as expected if $2<p<2.5$ in the
context of the standard afterglow model.

Excluding the UL events, of which only the upper limits of
$\beta_{\rm OX}$ are available, there is only one burst with
$\beta_{\rm OX}>1.25$. This is the peculiar GRB 060128, which is
sub-luminous, an X-ray flash, and associated with a supernova (Pian
et al. 2006), and its $\beta_{\rm OX}=1.76$. Campana et al. (2006)
claimed that its large first-day optical fluxes, behaving uniquely
and peaking near 11~hr, were powered by SN shock breakout rather
than a genuine GRB afterglow (see also Wang et al. 2007).

A significant fraction of both the OT events and UL events lies below
the $\beta_{\rm OX}=0.5$ criterion line, and hence are classified as dark
bursts ({\it red} symbols).  The true dark fraction could be
underestimated since some UL events, with an upper-limit $\beta_{\rm
OX}$ not much larger than 0.5, would have possibly turned out to be
dark had their optical flux densities been measured. The majority of
the dark bursts are long bursts and only three are short, while there
are a few more long events, but no short ones, in the ``gray'' region
($0.5<\beta_{\rm OX}<0.6$; {\it green} symbols).

%\begin{figure}[!hbp]

\subsection{Short bursts}\label{sect:ShortBursts}

In terms of their $\beta_{\rm OX}$ distribution, the difference
between the 19 short bursts and the long bursts is not significant,
as indicated by the K-S test $p$-value which is 0.4 for OT events
only or 0.2 if UL events are included. The optical afterglows of
short bursts as a whole are fainter than those of long bursts (Kann
et al. 2008), but so are their X-ray afterglows and prompt emissions
(Nysewander et al. 2009). Consequently, they occupy the lower-left
region of Fig. \ref{fig:XversusOptFlux} (see also Gehrels et al.
(2008) but for a much smaller sample). Compared with long bursts,
they are not preferentially optically dark relative to X-rays.

In our sample, three short bursts are dark, i.e., one OT (GRB
070809) and two UL events (GRB 061210 and GRB 070724A). Gehrels et
al. (2008), however, found no optically dark short bursts. We note
that GRB 061210 (the XRT observation which started 2 d later)
was not included in their sample, and neither was the optical
upper limit of GRB 070724A. The classification of GRB 070809 is
tricky since our $\beta_{\rm OX}$ value, 0.48, and theirs, 0.51,
are actually very similar, both bordering the $\beta_{\rm OX}=0.5$
separation line.

Although the $Swift$ satellite allows statistical studies on the
optical darkness of short bursts to be performed for the first time,
the current sample is still small. In the following sections, we
will focus on the optical darkness of long bursts. GRB 060614 and
GRB 060505 were also listed by us as long bursts despite the
on-going hot debates on their exact nature (e.g., Fynbo et al. 2006;
Zhang et al. 2007b).

\subsection{Long bursts}\label{sect:LongBurstsDis}

The total dark ratio of our $Swift$ long burst sample is $\gtrsim
12\%$, or possibly even $\gtrsim 18\%$ if all the gray UL events are
regarded as potential dark bursts. The ratio matches both the pre-$Swift$
result well, as concluded by Gehrels et al. (2008), and that
of our short bursts. Including the 10 high-$z$ events would not
change the conclusion (see \S \ref{sect:HighZ}). By comparison,
Cenko et al. (2009) reported a dark fraction as large as $\sim 50\%$
in the GRB sample that were followed up by the Palomar 60-inch
telescope, and Melandri et al. (2008) found that most of the GRBs
observed but undetected by the 2-m Liverpool and Faulkes telescopes
are dark bursts. Those samples are much smaller than ours and their
data inter-/extrapolation methods are somewhat different. The
23 Cenko GRBs before 2008 are all included in our sample, and
contribute only 4 dark bursts and 2 gray ones according to our
$\beta_{\rm OX}$ values. However, those authors only tabulated the
$\beta_{\rm OX}$ values at 1000~sec. As a test, we interpolated or
extrapolated their observed optical photometry to 11~hr and
recalculated $\beta_{\rm OX}$ and obtained only 3 dark bursts and 3
gray ones. Regarding the $Swift$ GRBs in Melandri et al. (2008), as
stated in Section \ref{sect:OptXray}, the $R$-band upper limit reported
there was adopted in our sample if it resulted in the best
available constraint on $\beta_{\rm OX}$.

\begin{figure}[]
\centering
   \includegraphics[bb=30 160 565 700,width=0.49\textwidth]{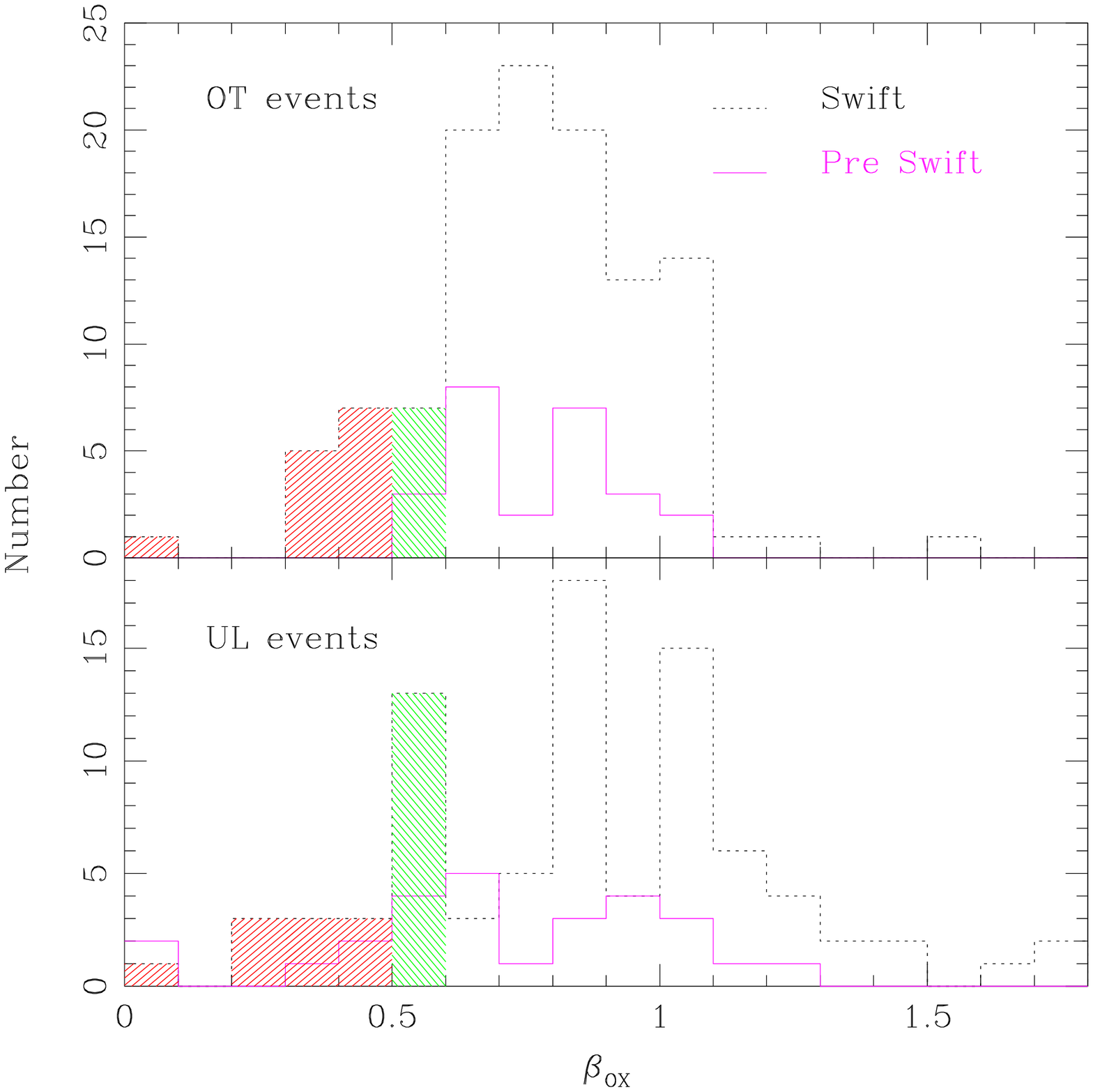}
   \includegraphics[bb=30 160 565 700,width=0.49\textwidth]{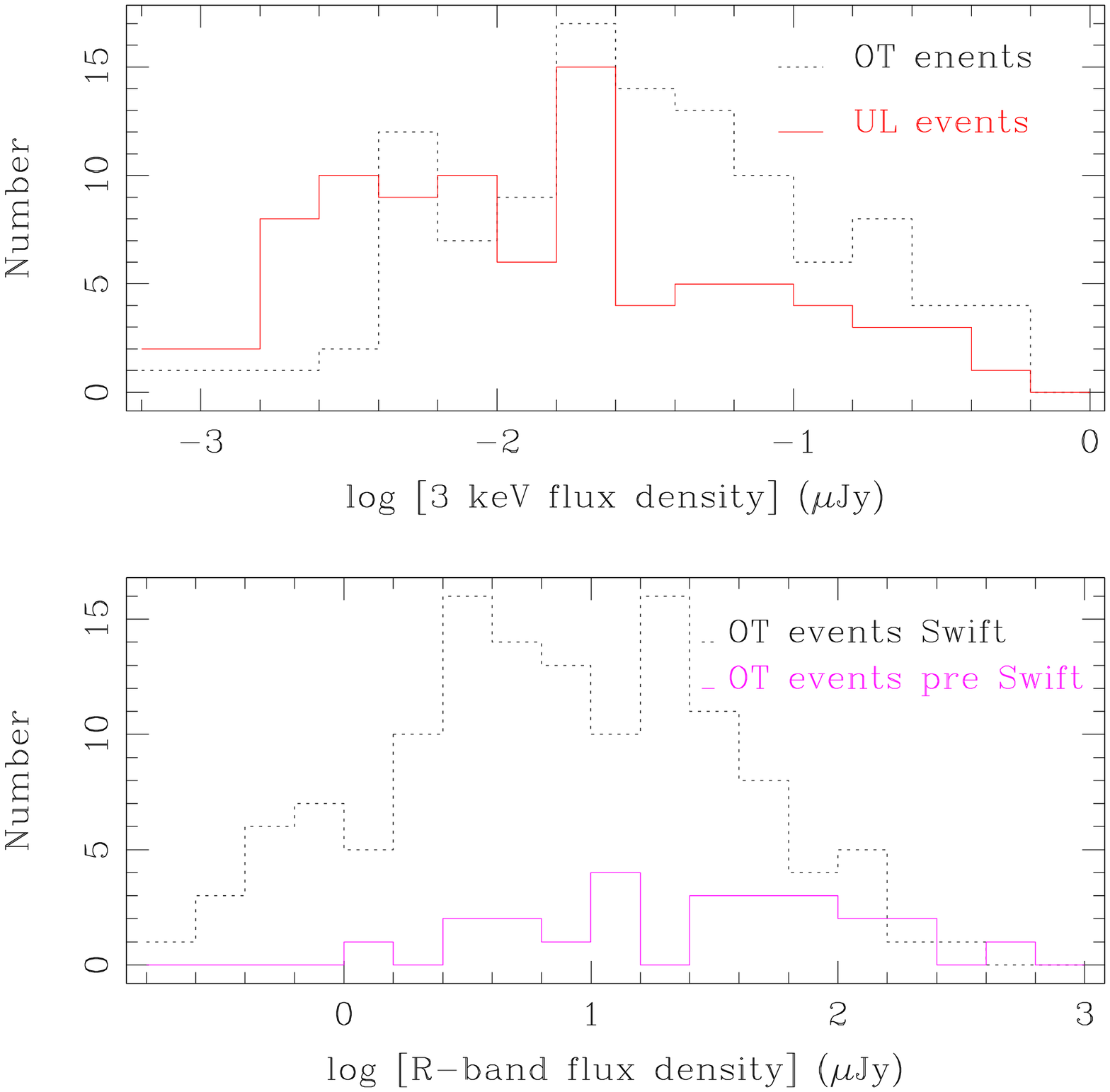}
   \caption{{\it Left}: $\beta_{OX}$ histogram distributions of the $Swift$
   ({\it dashed black}) and pre-$Swift$ ({\it solid pink}) long bursts, with OT events shown
   in the {\it upper} panel and UL events in the {\it lower} panel. The filled
   color of {\it red} indicates dark bursts, while {\it green} shows gray bursts.
   {\it Upper right}: Histogram distributions of the X-ray afterglow flux
   density at 3~keV and 11~hr of the $Swift$ OT ({\it dashed black}) and UL ({\it solid red})
   long bursts. {\it Lower right}: Histogram distribution of the $R$-band
   afterglow flux density at 11~hr of the $Swift$ OT long bursts ({\it dashed black})
   compared with that of the pre-$Swift$ ones ({\it solid pink}; Jakobsson et al. 2004).
   \label{fig:LongBurstsDis}}
\end{figure}

The {\it left} panel of Fig. \ref{fig:LongBurstsDis} shows the
$\beta_{\rm OX}$ histograms of all the long bursts of our sample ({\it
dashed black}) except for the high-$z$ ones, {\it upper} for the 113 OT
events and {\it lower} for the 87 UL ones. The distributions of the
Jakobsson et al. (2004) sample of 52 pre-$Swift$ GRBs ({\it solid
pink}) are also plotted for comparison.

The $Swift$ OT distribution is somewhat asymmetric, peaking around
0.8, with more below the peak value than above it. There are 13 dark
bursts ($\beta_{\rm OX}<0.5$) identified and 7 gray ones
($0.5<\beta_{\rm OX}<0.6$). The pre-$Swift$ sample is much smaller,
which may account for its narrower distribution, among which
Jakobsson et al. (2004) found no dark bursts. However, according to
the statistics, the two samples are not significantly different, with a
K-S test $p$-value as large as 0.9.

The $Swift$ UL distribution seems biased towards large $\beta_{\rm
OX}$ compared with the OT events (K-S test $p$-value as small as
0.004), not unexpected since their $\beta_{\rm OX}$ values are
simply also upper limits. However, one still finds 10 dark UL cases and
12 gray ones. Like the OT samples, the $Swift$ and pre-$Swift$ UL
samples are also not significantly different, with a K-S test
$p$-value of 0.2. Note that all the 5 dark bursts identified in the
pre-$Swift$ sample were UL events.

As shown in the {\it upper right} panel of Fig. \ref{fig:LongBurstsDis},
the 11-hr 3-keV X-ray flux densities of the $Swift$ UL events are,
on average, lower than the OT ones (K-S test $p$-value of $4\times 10^{-4}$).
The mean value of the former is $\sim 10^{-2.0\pm 0.8}$~$\mu$Jy, while
the latter is $\sim 10^{-1.5\pm 0.6}$~$\mu$Jy. We have also found a similar
trend regarding the BAT fluence of the two samples, although the statistical
significance is not as high (K-S test $p$-value of 0.04). On the one
hand, weaker gamma-ray bursts also tend to release less energy
during the afterglow stage, both in X-rays and in optical photons.
On the other hand, optical afterglow detections rely on quick and
accurate BAT and XRT localizations to become efficient, the latter
clearly being favored by large BAT fluence and large X-ray flux
densities.

The optical afterglows of the $Swift$ OT events are on average much
fainter than the pre-$Swift$ ones of Jakobsson et al. (2004),
demonstrating not only the ever-improving ensemble of GRB-dedicated
optical facilities but also the excellent synergy between $Swift$
detections and ground-based follow-up observations. The respective
11-hr $R$-band flux density distribution plotted in the {\it lower
right} panel of Fig. \ref{fig:LongBurstsDis} has a mean value of
$\sim 0.03\pm 0.1$~mJy for $Swift$ and of $\sim 0.2\pm 0.8$~mJy for
pre-$Swift$. The two distribution are different with a K-S test
$p$-value as small as 0.008. The corresponding distributions of the
UL events are however indistinguishable (K-S test $p$-value of 0.8).

%\begin{figure}[!hbp]

\section{Discussions on the Nature of Dark Bursts}
\label{sect:Discussions}
\subsection{The high-redshift scenario}\label{sect:HighZ}

We have singled out the 10 GRBs that have measured redshifts
approximately higher than 4 (see Table \ref{tab:HighZEvents}), in
order to check if some bursts are optically dark simply due to very
high redshifts. For $z\gtrsim 4$, the observed $R$-band flux is
greatly depressed by the strong intergalactic or interstellar H
Lyman absorptions at $\lambda_{\rm obs}\leq 1025(1+z){\rm \AA}$
(Lamb \& Reichart 2000). Two values of $\beta_{\rm OX}$ were
calculated for each burst among the high-$z$ sample. The first,
$\beta_{\rm OX}1$, was obtained directly from the observed $R$-band
flux, while to get $\beta_{\rm OX}2$ we extrapolated the observed
NIR or IR flux (unaffected by Lyman absorptions for the given
redshift) to the $R$ band adopting a spectral index of $1$.

It can be shown that the high-$z$ H Lyman absorption does contribute
to a few dark bursts, but it is unlikely to be a major factor for the $Swift$
GRB sample. Taking the $\beta_{\rm OX}1$ values, the high-$z$
sub-sample members seem somewhat ``darker'' than those at lower redshifts.
There are two events with $\beta_{\rm OX}1\leq 0.5$ and three with
$0.5<\beta_{\rm OX}1\leq 0.6$. On the other hand, the $\beta_{\rm
OX}2$ values (i.e., having been de facto corrected for H Lyman
absorptions) decrease the number of high-$z$ dark bursts and those
with gray ones both to 1. However, high-$z$ GRBs identified so far are
rare. Among the GRBs detected by $Swift$ and with measured redshifts,
only a small fraction are located  at $z\gtrsim 4$, or $\sim 10\%$
up to the end of the year 2007. However, we also sound a note of
caution that, among our whole long GRB sample, 9 of the 13 dark OT
events, 9 of the 11 dark UL pmes, and 10 of the 12 gray UL ones have
no measured redshifts. Some would turn out to be high-$z$ events if
their redshifts were measured.

\subsection{Dust extinction}\label{sect:Extinction}

%\begin{figure}[!hbp]
\begin{figure}[]
\centering
   \includegraphics[bb=30 160 570 580,width=0.49\textwidth]{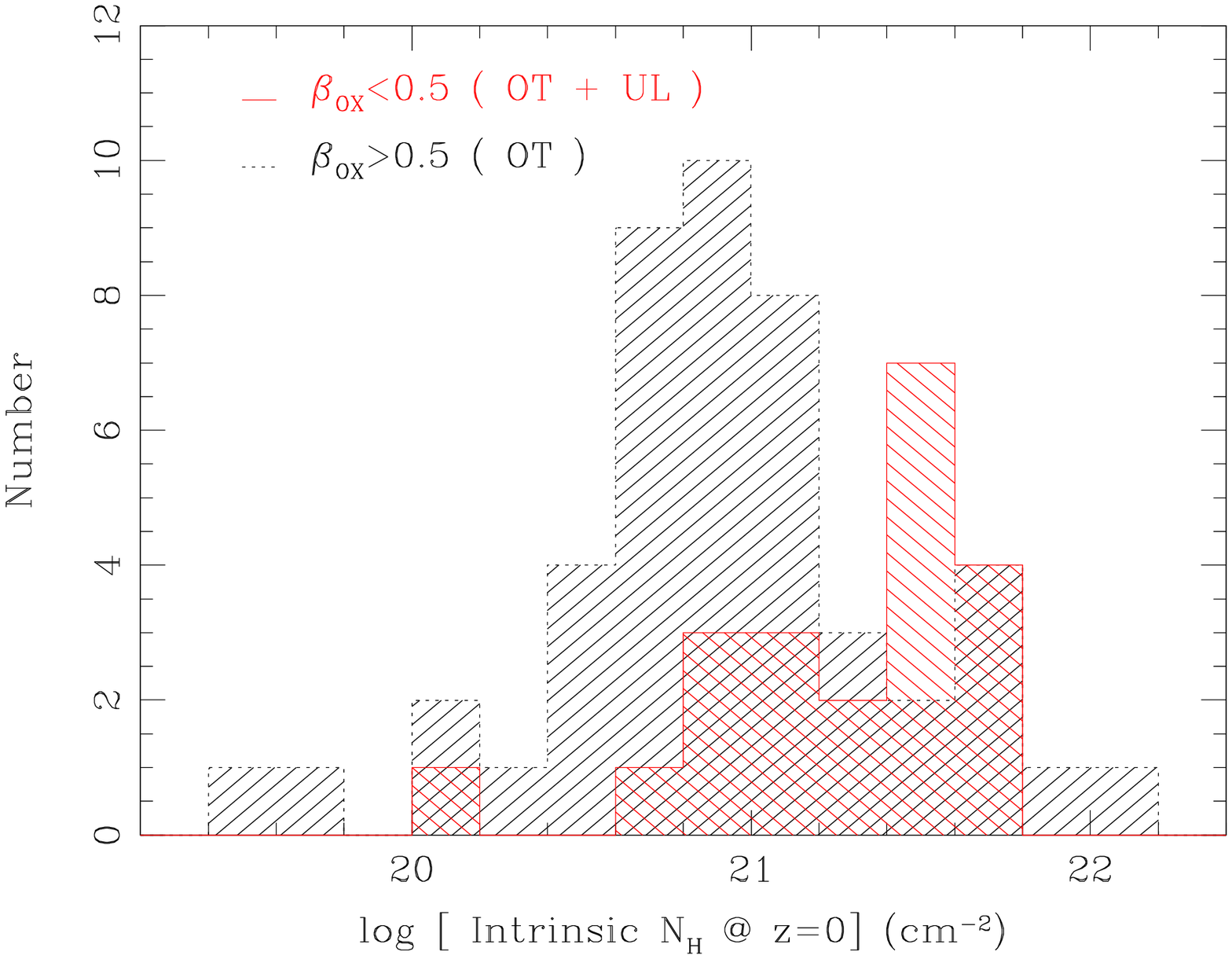}
   \includegraphics[bb=30 160 570 580,width=0.49\textwidth]{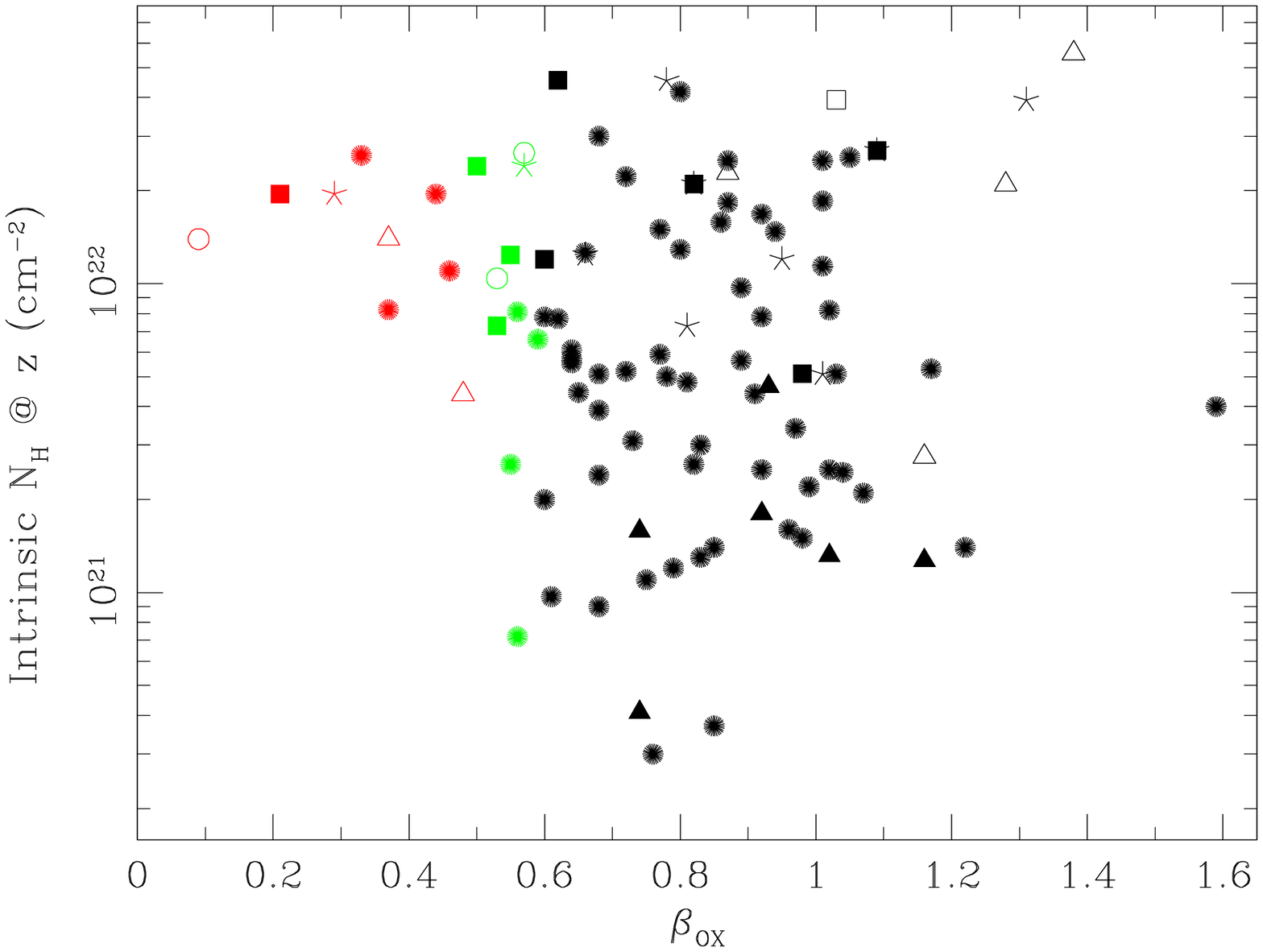}
   \caption{{\it Left}: Histogram distribution of the intrinsic hydrogen
column density $N^0_{\rm H}$ (assuming $z=0$), in excess of the Galactic value,
of dark bursts ({\it solid red};  both OT and UL events included) and normal ones
({\it dashed black}; OT only). {\it Right}: The same as in the left but for
the value measured in the host galaxy's rest-frame $N^z_{\rm H}$. The meanings
of the symbols and colors are the same as in Fig. \ref{fig:XversusOptFlux}.
\label{fig:NH}}
\end{figure}

It seems natural to hypothesize that dust extinction may account for
at least some fraction of dark bursts, in view of the compelling
evidence for the connection of long GRBs with the deaths of massive
stars and with star-forming regions (see Woosley \& Bloom 2006 and
references therein). This scenario was first proposed by Groot et
al. (1998) and Paczy\'{n}ski (1998) to explain the failed optical
detection of GRB 970828, which can also be regarded as a
prototypical dark burst by the $\beta_{\rm OX}<0.5$ criterion of
Jakobsson et al. (2004). However, as pointed out by Lazzati et al.
(2002) and De Pasquale et al. (2003), it is complicated by
theoretical predictions of possible dust destruction by strong
GRB X-rays and/or UV flashes along the line of sight in the
circumburst environment (e.g., Waxman \& Draine 2000; Fruchter et
al. 2001). On the other hand, although still rare, highly
extinguished GRBs were discovered in multi-band optical/IR
observations, some of which are dark bursts according to $\beta_{\rm
OX}$ (e.g., Rol et al. 2007; Tanvir et al. 2008; Jaunsen et al. 2008)
while others seem not to be.

We used the hydrogen column density $N_{\rm H}$ as a proxy for the
strength of local dust extinction to make statistical comparisons
between dark bursts and normal ones, since the number of GRBs with
known optical extinction values is far too low for such a
purpose. The $N_{\rm H}(10^{22}~{\rm cm}^{-2})/A_{V}$ value were
estimated to be 0.18, 0.7, and 1.6 for the Milky Way, LMC, and SMC,
respectively (see Schady et al. 2007 and references therein),
although the values derived from GRB afterglow observations were, in
general, much smaller possibly suggesting a lower dust-to-gas ratio
in the GRB local environment (Galama \& Wijers 2001; Stratta et al.
2004; Kann et al. 2006; Starling et al. 2007; but see Schady et al.
2007).

The $N^0_{\rm H}$ values for dark bursts are on average higher than
for normal ones ({\it left} panel; Fig. \ref{fig:NH}). These are the
X-ray absorption column densities in excess of the Galactic values,
assuming zero redshift for the host galaxy (Evans et al. 2007).
Neither short bursts nor high-redshift ones ($z\gtrsim 4$) are
included. For dark bursts ({\it solid red}) both OT and UT events
are included, while for normal ones ({\it dashed black}) only OT
events are counted since some UL events of $\beta_{\rm OX}>0.5$
could actually be dark. In total, 21 dark and 47 normal bursts
are plotted in the figure. All but 5 of the dark bursts have
$N^0_{\rm H}>10^{21}$~cm$^{-2}$, corresponding to a ratio of
$\sim 76\%$, while the same high-$N^0_{\rm H}$ ratio for normal
ones is only $\sim 40\%$. Applying a K-S test, we found that
the probability that dark bursts and normal ones are drawn from the
same $N^0_{\rm H}$ distribution is very small, only $\sim 0.4\%$.
The histogram is only for long bursts. However, the conclusion is not
changed by including short bursts, whose numbers are very small.
Lin (2006) claimed somewhat different results, but his dark bursts
just meant no-detection by the UVOT and in total only 25 $Swift$
GRBs were studied.

A more physically reasonable comparison can be made using the
intrinsic $N_{\rm H}$ values in the rest frame of the GRB host
galaxy, i.e., $N^z_{\rm H}$. Dark bursts are clearly inclined to
have large $N^z_{\rm H}$ values despite possibly small statistics.
This is shown in Fig. \ref{fig:NH} by an $N^z_{\rm H}-\beta_{\rm OX}$
plot ({\it right} panel), instead of a histogram since the number of
dark bursts whose $N^z_{\rm H}$ values are available is too small. Both
long ({\it circles}) and short bursts ({\it triangles}) are
included, as well as high-redshift ones ({\it squares} and {\it
asterisks}). All the 9 dark bursts plotted here ({\it red}) have
$N^z_{\rm H}>4\times 10^{21}$~cm$^{-2}$, as do most of the gray ones
({\it green}). In contrast, the $N^z_{\rm H}$ value of about half of
the normal events ({\it black}) is well below $4\times
10^{21}$~cm$^{-2}$.

For further evidence of dust extinction dominating the formation of
dark bursts as suggested by their very large $N_{\rm H}$, we looked
to the literature for any measured optical extinction values of the
dark bursts in our sample. Our search, though by no means
comprehensive, resulted in 4 bursts, which are GRB 050401
($A_V\approx 0.6$; Watson et a. 2006), GRB 060923A ($A_V\approx 3$;
Tanvir et al. 2008), GRB 070306 ($A_V\approx 4.9$; Jaunsen et al.
2008), and GRB 070802 ($A_V\approx 0.8-1.5$; El\'{i}asd\'{o}ttir et al.
2009; see also Kr\"{u}hler et al. 2008). Remarkably, all of them
experienced strong dust extinction. Moreover, after correcting for
their local optical extinction values, all had $\beta_{\rm
OX}>0.5$, i.e., they were no longer optically dark! Note the X-ray flux
densities that we compiled have already been corrected for gas
absorptions because of the nature of the X-ray data reduction procedures
(Evans et al. 2007).

Systematic rapid and deep afterglow observations simultaneously
taken in many optical and NIR bands are required to reliably
constrain the local dust-extinction characteristics for dark and
normal bursts. The $A_V$ values cited above were derived assuming
one of the Milky Way, LMC, and SMC extinction laws, while the true
cases in the GRB environments or host galaxies can be very different
(Li et al. 2008). Such an observational campaign is already being
realized in the ongoing GROND project (e.g., Kr\"{u}hler et al. 2008),
as well as being proposed for future GRB space missions like SVOM
(Basa et al. 2008), JANUS (Roming et al. 2008), and EXIST (Grindlay
et al. 2009).

\subsection{Non-standard emission mechanisms}

Our studies do not exclude the possibility that some dark bursts are
attributable to the intrinsic mechanism by which the GRB afterglow
emissions are produced. As shown in the {\it left} panel of Fig.
\ref{fig:NH}, one long dark burst does have a very low $N^0_{\rm H}$
value. This is GRB 060510A with $N^0_{\rm H}=1.1\times
10^{20}$~cm$^{-2}$, corresponding to marginal apparent dust
extinction in the Milky Way, LMC, or SMC $N_{\rm H}/A_V$
relationships. The redshift of the GRB is unknown. Of course, the
intrinsic column density $N^z_{\rm H}$ could be much higher if it took
place at a relatively high redshift.  What is not included in the
figure is the short dark GRB 070809 with $N^0_{\rm H}=1.2\times
10^{20}$~cm$^{-2}$. Being a short GRB, it was very likely lying at
$z<1$ and hence $N^z_{\rm H}$ is also not expected to be large.
Urata et al. (2007) also argued that the optical darkness of GRB
051028, which was detected by the HETE-2 satellite, cannot be
explained by local dust extinction since the X-ray fitting $N_{\rm
H}$ is consistent with the Galactic value.

%\begin{figure}[!hbp]
\begin{figure}[]
\centering
   \includegraphics[bb=25 160 565 700,width=0.7\textwidth]{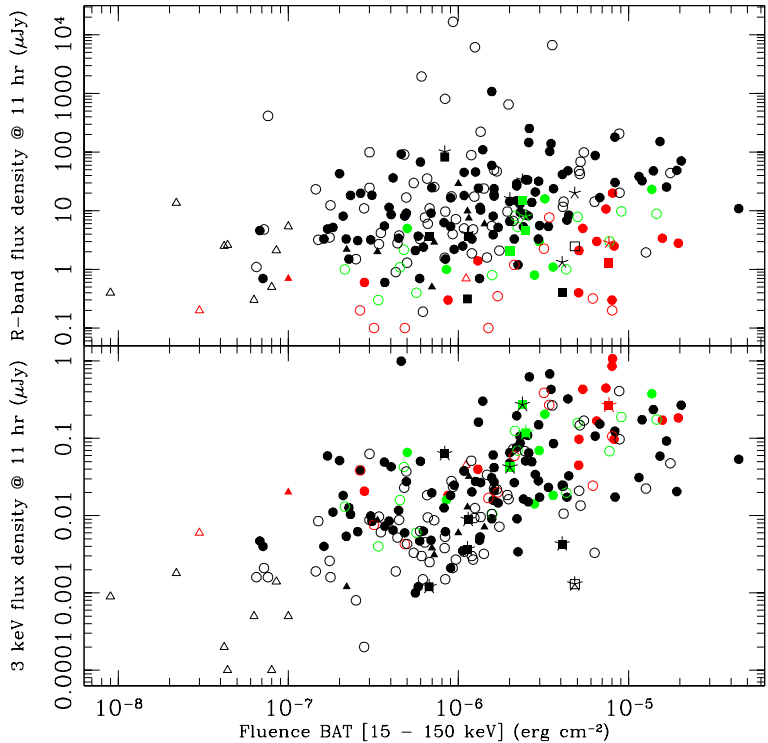}
   \caption{$R$-band optical ({{\it upper} panel}) and 3 keV X-ray
({\it lower} panel) flux density at 11 hr as a function of the $Swift$
BAT fluence. The meanings of the symbols and colors are the same as
in Fig. \ref{fig:XversusOptFlux}. \label{fig:FluFx}}
\end{figure}

The dark burst criterion adopted by us is defined in the context of
the standard afterglow model. It assumes that both the X-ray and
optical afterglows are just different segments of the same
synchrotron radiation spectrum of relativistic electrons that are
accelerated in the forward shock when the GRB ejecta collide with
circumambient material (e.g., Sari et al. 1998). This paradigm has
been seriously challenged in the $Swift$ era by several
observational facts (see Zhang 2007a and references therein), in
particular, the occurrence of an early shallow-decay phase in many
X-ray afterglow light curves (e.g., Liang et al. 2007), which was
not often accompanied by similar behavior in the optical band, and
the remarkable paucity of a highly-expected achromatic LC jet break
(e.g., Liang et al. 2008). Many remedies have been proposed,
including time-dependent microphysical parameters (e.g., Fan \&
Piran 2006), the contribution by long-lived reverse shocks (e.g.,
Genet et al. 2007), X-rays dominated by ``late prompt emission''
(Ghisellini et al. 2007), the importance of inverse-Compton scattering
(Panaitescu 2008), scattering of X-rays by dust (Shao \& Dai 2007),
etc. The X-ray and optical afterglows may result from different
emission mechanisms or regions, e.g., from relativistic forward
and reverse shocks respectively. This forward-reverse-shock model
was proposed early on by Dai (2004) and then numerically studied by
Yu \& Dai (2007) (see also Genet et al. 2007). It may be the emission
from one of the two shocks that leads to optically dark GRBs.
In addition, for the pre-$Swift$ dark GRB 001025A, Pedersen et
al. (2006) invoked an extra thermal component to fit the X-ray
spectrum which also solves its optical-darkness puzzle. Also, the
afterglow emissions in the case of an electron energy spectral index
$p<2$, which can naturally lead to $\beta_{\rm OX}<0.5$, were
calculated by Dai \& Cheng (2001), although it is difficult to make
compatible with general acceleration mechanisms.

Finally, as shown in Fig. \ref{fig:FluFx}, dark and normal bursts
follow a similar correlation with the X-ray flux density and
$\gamma$-ray fluence ({\it lower}; see also Gehrels et al. 2008),
while dark bursts lie considerably below normal ones in the
optical-$\gamma$ diagram ({\it upper}). On the one hand, this
indicates that early X-ray abnormalities like flares and the shallow
decay do not increase the apparent ratio of dark bursts. On the
other hand, this may further support the dust-extinction hypothesis
for dark bursts since they are clearly dim in the optical band rather
than bright in X-rays.

\section{Comparisons with Other Studies}\label{sect:Comparison}

We have compiled a large sample for statistical studies of the optical
darkness of GRBs. The data set is uniform in the sense that all
the GRBs were detected by the same instrument, i.e. the $Swift$ BAT.
Gehrels et al. (2008) have calculated $\beta_{\rm OX}$ values or upper
limits for a much smaller $Swift$ sample, consisting of only 41 long bursts and
10 short ones. On the one hand, they did not search for the GRBs detected
in the last 5 months of 2007 which we have done. On the other hand,
they only considered GRBs having optical measurements within a factor
of 2 of 11~hr. Probably as a consequence, we have identified 3 dark short
bursts but they found none. Notwithstanding, the dark ratio among long
bursts is similar between the two studies and is also consistent with
the pre-$Swift$ sample (De Pasquale et al. 2003; Jakobsson et al. 2004).
In contrast, both Melandri et al.(2008) and Cenko et al.(2009) claimed
a dark-burst occurrence as high as $\sim 50\%$ among the few dozen
GRBs that have been observed with their specific optical telescopes.
However, the light curve slopes adopted there for optical and X-ray data
inter-/extrapolation are different from ours. Moreover, the $\beta_{\rm OX}$
in Cenko et al. (2009) was defined at $10^3$~s. Nevertheless, their
optical measurements have been taken into account when compiling our
data set, but with necessary modifications to meet our data definitions.

There are other optical-darkness statistics performed on $Swift$ GRBs but using
optical non-detection as the dark-burst definition, e.g., Lin (2006) and
Bal\'{a}zs et al. (2008). The former found no systematic difference
in X-ray $N_{\rm H}$ between their ``dark'' and ``bright'' groups, totaling
25 GRBs, while the latter came to a contradictory conclusion for a much larger
sample. We cannot make direct comparisons between these results and ours.
Nardini et al. (2008) compiled the $R$-band upper limits of optically
non-detected $Swift$ GRBs to argue that most belong to their
``underluminous optical'' family. More recently, after the paper had
been submitted, Perley et al. (2009) reported host galaxy studies
of 14 ``dark'' GRBs, mixing optical non-detection and the $\beta_{\rm OX}<0.5$
criterion in order to get a decent sample size. Like us, they
identified dust extinction as the main cause of dark bursts. Finally,
van der Horst et al. (2009) proposed a new definition for dark bursts,
i.e., $\beta_{\rm OX}<\beta_{\rm X}-0.5$ where $\beta_{\rm X}$ is
the X-ray spectral index. By doing so, they can exclude the potential
cases of $p<2$ being classified as optically dark.

\begin{acknowledgements}
We thank J. Hu, Y. Qiu, L. Xin, and J. Wei for stimulating
discussions. This work was supported by the National Natural Science
Foundation of China (Grant Nos. 10673014 and 10873017) and by the
National Basic Research Program of China (Grant No. 2009CB824800).
This research has made use of data supplied by the UK $Swift$
Science Data Center at the University of Leicester.
\end{acknowledgements}

\clearpage

\begin{deluxetable}{lcccccccccc}
 \tabcolsep 0.5mm
 %\tabletypesize{\scriptsize}
 %\tablecolumns{9}
 \tablewidth{0pt}
 %\tablenum{1}
 \tablecaption{$Swift$ GRB sample catalog\label{tab:SwiftBursts}}
  \tablehead{
\colhead{ID}  &  \colhead{Redshift}\tablenotemark{\dag} &  \colhead{BAT Fluence\tablenotemark{\ddag}}  &  \colhead{Flux Density} &  \colhead{Flux Density} &  \colhead{$\beta_{\rm OX}$}   & \colhead{Intrinsic $N_{\rm H}$\tablenotemark{\oplus}} & \colhead{Intrinsic $N_{\rm H}$\tablenotemark{\oplus}}\\
\colhead{}  &  \colhead{$z$}  &  \colhead{[15-150 keV]} &  \colhead{$R$ @ 11h}  &  \colhead{3~keV @ 11h}  &  \colhead{@ 11h } & \colhead{@ $z=0$} & \colhead{@ GRB $z$}\\
\colhead{}  &  \colhead{}  &  \colhead{10$^{-7}$ erg cm$^{-2}$} &
\colhead{$\mu$Jy} &  \colhead{10$^{-3}$ $\mu$Jy} &  \colhead{}  &
\colhead{10$^{21}$ cm$^{-2}$} & \colhead{10$^{21}$ cm$^{-2}$} }
 \startdata
& & & Short & bursts & & &\\
\hline
\\
OT& & & & & & &\\
------------& & & & & & &\\
\\
$\beta_{\rm OX} <$ 0.5 & & & & & & &\\
------& & & & & & &\\
\\
070809   &           &  1.0     &  0.7      &  20.2    &  0.48  &  0.12   &               \\
\\
$\beta_{\rm OX} \geq$ 0.6 & & & & & & &\\
------& & & & & & &\\
\\
050724   &  0.258    &  9.98    &  29.1     &  6.0     &  1.16  &              &  1.27$^{a}$    \\
051221A  &  0.547    &  11.50   &  7.5      &  32.0    &  0.74  &              &  1.58$^{a}$    \\
051227   &           &  6.99    &  0.5      &  3.8     &  0.66  &  0.79   &               \\
060313   &           &  11.30   &  9.9      &  12.9    &  0.90  &              &               \\
061006   &  0.4377   &  14.20   &  6.0      &  7.1     &  0.92  &              &  1.8$^{a}$     \\
061201   &  0.111    &  3.34    &  2.0      &  8.7     &  0.74  &  0.33   &  0.41$^{\otimes}$    \\
070714B  &  0.92     &  7.2     &  2.9      &  3.1     &  0.93  &              &  4.65$^{a}$    \\
071227   &  0.383    &  2.2     &  2.2      &  1.2     &  1.02  &  0.57   &  1.32$^{\otimes}$    \\
\\
UL& & & & & & &\\
------------& & & & & & &\\
\\
$\beta_{\rm OX} <$ 0.5 & & & & & & &\\
------& & & & & & &\\
\\
061210   &  0.4095   &  11.10   &  0.7      &  45.1    &  0.37  &  6.8    &  13.98$^{\otimes}$   \\
070724A  &  0.457    &  0.3     &  0.2      &  6.0     &  0.48  &  1.99   &  4.39$^{\otimes}$    \\
\\
$\beta_{\rm OX} \geq$ 0.6 & & & & & & &\\
------& & & & & & &\\
\\
050509B  &  0.226    &  0.09    &  0.4      &  0.9     &  0.83  &              &  0.0018$^{a}$  \\
050813   &  1.8      &  0.44    &  2.6      &  0.1     &  1.38  &  6.4    &  55.62$^{\otimes}$   \\
051210   &           &  0.85    &  2.1      &  1.4     &  1.00  &  2.5    &               \\
060801   &           &  0.80    &  0.5      &  0.1     &  1.16  &              &  2.75$^{a}$    \\
061217   &  0.827    &  0.42    &  2.5      &  0.2     &  1.28  &  5.9    &  20.92$^{\otimes}$   \\
070209   &           &  0.22    &  13.6     &  1.8     &  1.22  &              &               \\
070429B  &  0.904    &  0.63    &  0.3      &  0.5     &  0.87  &  5.9    &  22.81$^{\otimes}$   \\
070729   &           &  1.0     &  5.4      &  0.5     &  1.26  &  0.39   &               \\
\\
\hline
& & & Long & bursts & & &\\
\hline
\\
OT& & & & & & &\\
------------& & & & & & &\\
\\
$\beta_{\rm OX} <$ 0.5 & & & & & & &\\
------& & & & & & &\\
\\
050219B  &           &  158.00  &  3.4      &  172.2   &  0.41  &  1.24   &               \\
050401   &  2.90     &  82.20   &  2.5      &  97.2    &  0.44  &              &  19.5$^{b}$    \\
050713A  &           &  51.10   &  2.1      &  97.3    &  0.42  &  2.61   &               \\
051008   &           &  50.90   &  0.4      &  44.9    &  0.30  &  3.15   &               \\
060510A  &           &  80.50   &  20.0     &  1072.1  &  0.40  &  0.11   &               \\
060923A  &           &  8.69    &  0.3      &  18.3    &  0.38  &  0.96   &               \\
061222A  &           &  79.90   &  0.3      &  858.2   &  -0.1  &  2.59   &               \\
070306   &  1.4959   &  53.80   &  5.0      &  431.7   &  0.33  &  3.34   &  26.0$^{e}$    \\
070419B  &           &  73.60   &  10.7     &  446.7   &  0.43  &  0.57   &               \\
070508   &  0.82     &  196.00  &  2.8      &  183.7   &  0.37  &  2.34   &  8.23$^{\otimes}$    \\
070802   &  2.45     &  2.8     &  0.6      &  20.7    &  0.46  &              &  11.0$^{a}$    \\
071021   &           &  13      &  1.4      &  39.8    &  0.48  &  1.37   &               \\
071025   &           &  65      &  3.0      &  167.8   &  0.39  &  0.82   &               \\
\\
0.5 $\leq \beta_{\rm OX} <$ 0.6 & & & & & & &\\
------& & & & & & &\\
\\
050315   &  1.949    &  32.20   &  16.0     &  205.2   &  0.59  &              &  6.6$^{b}$     \\
050915A  &           &  8.50    &  1.0      &  16.0    &  0.56  &  0.89   &               \\
060908   &  2.43     &  28.00   &  0.8      &  14.2    &  0.55  &              &  2.6$^{a}$     \\
061121   &  1.314    &  137.00  &  23.2     &  377.8   &  0.56  &              &  8.1$^{b}$     \\
070129   &           &  29.80   &  3.0      &  69.5    &  0.51  &  0.498  &               \\
070721B  &  3.626    &  36      &  1.1      &  18.3    &  0.56  &  0.029  &  0.72$^{\otimes}$    \\
071118   &           &  5.0     &  5.0      &  65.6    &  0.59  &  1.55   &               \\
\\
$\beta_{\rm OX} \geq$ 0.6 & & & & & & &\\
------& & & & & & &\\
\\
041223   &           &  167.00  &  25.4     &  92.4    &  0.76  &  0.31   &               \\
050126   &  1.29     &  8.38    &  1.5      &  6.2     &  0.75  &              &  1.1$^{a}$     \\
050215B  &           &  2.27    &  1.5      &  12.7    &  0.65  &              &               \\
050306   &           &  115.00  &  38.5     &  31.1    &  0.97  &              &               \\
050318   &  1.44     &  10.80   &  45.2     &  37.9    &  0.96  &              &  1.6$^{b}$     \\
050319   &  3.24     &  13.10   &  24.3     &  162.1   &  0.68  &              &  0.9$^{a}$     \\
050406   &  2.44     &  0.68    &  4.6      &  4.7     &  0.94  &  1.1    &  14.73$^{\otimes}$   \\
050416A  &  0.6535   &  3.67    &  5.5      &  49.3    &  0.64  &              &  6.1$^{b}$     \\
050525A  &  0.606    &  153.00  &  151.9    &  58.9    &  1.07  &              &  2.1$^{b}$     \\
050603   &  2.821    &  63.60   &  87.6     &  107.0   &  0.91  &              &  4.4$^{a}$     \\
050607   &           &  5.92    &  0.7      &  4.7     &  0.68  &  0.6    &               \\
050712   &           &  10.80   &  18.2     &  37.8    &  0.84  &  0.4    &               \\
050721   &           &  36.20   &  23.8     &  85.7    &  0.77  &  1.25   &               \\
050801   &  1.56     &  3.10    &  18.4     &  8.4     &  1.05  &  5.1    &  25.60$^{\otimes}$   \\
050802   &  1.71     &  20.00   &  14.4     &  65.3    &  0.73  &  0.263  &  3.1$^{b}$     \\
050820A  &  2.612    &  34.40   &  103.2    &  681.8   &  0.68  &  0.262  &  3.89$^{\otimes}$    \\
050824   &  0.83     &  2.66    &  19.9     &  38.3    &  0.85  &              &  0.37$^{a}$    \\
050826   &  0.297    &  4.13    &  36.4     &  6.5     &  1.17  &              &  5.3$^{b}$     \\
050908   &  3.350    &  4.83    &  8.1      &  6.0     &  0.98  &  0.21$^{c}$   &               \\
050915B  &           &  33.80   &  5.4      &  23.2    &  0.74  &  1.8    &               \\
050922C  &  2.198    &  16.20   &  23.9     &  27.2    &  0.92  &              &  2.5$^{a}$     \\
051006   &           &  13.40   &  11.4     &  5.3     &  1.04  &  5.46   &               \\
051016B  &  0.9364   &  1.70    &  4.9      &  59.2    &  0.60  &              &  7.8$^{b}$     \\
051109A  &  2.346    &  22.00   &  29.0     &  196.2   &  0.68  &              &  5.1$^{a}$     \\
051109B  &  0.080    &  2.56    &  3.1      &  6.2     &  0.85  &  0.98   &  1.4$^{b}$     \\
051111   &  1.55     &  40.80   &  41.6     &  24.8    &  1.01  &              &  18.5$^{b}$    \\
051117A  &           &  43.40   &  6.9      &  17.3    &  0.82  &  0.81   &               \\
060108   &  2.03     &  3.69    &  0.6      &  7.3     &  0.60  &              &  2$^{a}$       \\
060110   &           &  15.70   &  59.4     &  30.9    &  1.03  &  0.41   &               \\
060111A  &           &  12.00   &  3.3      &  20.3    &  0.69  &  1.32   &               \\
060111B  &           &  16.00   &  3.9      &  15.8    &  0.75  &  2.2    &               \\
060115   &  3.53     &  17.10   &  5.2      &  14.6    &  0.80  &              &  12.9$^{b}$    \\
060116$^{\tablenotemark{\circledast}}$   &           &  24.10   &  5.34     &  16.4    &  0.79  &  5.74   &               \\
060124   &  2.296    &  4.61    &  92.8     &  997.2   &  0.62  &              &  7.7$^{b}$     \\
060202   &  0.783    &  21.30   &  3.3      &  26.7    &  0.66  &  3.74   &  12.60$^{\otimes}$   \\
060203   &           &  8.98    &  5.4      &  18.2    &  0.77  &  0.71   &               \\
060204B  &           &  29.50   &  3.3      &  28.2    &  0.65  &  1.16   &               \\
060218   &  0.0331   &  15.70   &  1082.2   &  9.1     &  1.59  &              &  4.0$^{b}$     \\
060323   &           &  6.22    &  2.4      &  6.3     &  0.81  &  0.041  &               \\
060418   &  1.490    &  83.30   &  30.4     &  17.4    &  1.02  &              &  8.2$^{b}$     \\
060428A  &           &  13.90   &  110.0    &  302.8   &  0.80  &  1.33   &               \\
060428B  &           &  8.23    &  6.3      &  12.6    &  0.85  &  0.62   &               \\
060502A  &  1.51     &  23.10   &  9.4      &  86.7    &  0.64  &              &  5.7$^{b}$     \\
060505   &  0.089    &  9.44    &  2.2      &  24.8    &  0.61  &  0.81   &  0.97$^{\otimes}$    \\
060507   &           &  44.50   &  8.4      &  32.7    &  0.76  &  0.48   &               \\
060512   &  0.4428   &  2.32    &  18.3     &  10.4    &  1.02  &              &  2.5$^{b}$     \\
060526   &  3.221    &  12.60   &  45.9     &  27.7    &  1.01  &              &  11.4$^{b}$    \\
060604   &  2.68     &  4.02    &  8.6      &  43.0    &  0.72  &              &  22.2$^{b}$    \\
060605   &  3.78     &  6.97    &  24.8     &  19.7    &  0.97  &              &  3.4$^{a}$     \\
060607A  &  3.082    &  25.50   &  33.6     &  109.3   &  0.78  &              &  5.0$^{b}$     \\
060614   &  0.125    &  204.00  &  70.8     &  268.5   &  0.76  &  0.143  &  0.3$^{b}$     \\
060707   &  3.425    &  16.00   &  8.1      &  60.9    &  0.67  &              &               \\
060708   &           &  4.94    &  9.1      &  27.6    &  0.79  &  0.31   &               \\
060714   &  2.711    &  28.30   &  20.9     &  34.1    &  0.87  &              &  18.3$^{b}$    \\
060729   &  0.54     &  26.10   &  253.3    &  623.7   &  0.82  &              &  2.6$^{b}$     \\
060804   &           &  5.98    &  67.9     &  50.4    &  0.98  &              &               \\
060904B  &  0.703    &  16.20   &  17.8     &  21.0    &  0.92  &              &  7.8$^{b}$     \\
060906   &  3.686    &  22.10   &  14.7     &  9.1     &  1.01  &              &  25.0$^{a}$    \\
060912A  &  0.937    &  13.50   &  10.3     &  26.8    &  0.81  &              &  4.8$^{b}$     \\
060926   &  3.208    &  2.19    &  3.3      &  5.4     &  0.87  &              &  25.0$^{a}$    \\
061007   &  1.261    &  444.00  &  10.9     &  53.6    &  0.72  &              &  5.2$^{b}$     \\
061019   &           &  25.90   &  146.4    &  64.7    &  1.05  &  6.9    &               \\
061021   &           &  29.60   &  31.9     &  149.8   &  0.73  &  0.443  &               \\
061110A  &  0.758    &  10.60   &  2.5      &  3.5     &  0.89  &              &  9.7$^{b}$     \\
061110B  &  3.44     &  13.30   &  0.7      &  4.8     &  0.68  &              &  30.0$^{a}$    \\
061126   &  1.1588   &  67.70   &  16.8     &  153.0   &  0.64  &  1.11   &  5.54$^{\otimes}$    \\
061222B  &  3.355    &  22.40   &  1.2      &  3.4     &  0.80  &              &  41.8$^{b}$    \\
070110   &  2.352    &  16.20   &  18.9     &  41.9    &  0.83  &              &  1.3$^{d}$     \\
070208   &  1.165    &  4.45    &  3.4      &  11.7    &  0.77  &              &  5.9$^{a}$     \\
070224   &           &  3.05    &  3.2      &  9.9     &  0.79  &              &               \\
070318   &  0.836    &  24.80   &  34.0     &  51.0    &  0.89  &              &  5.64$^{a}$    \\
070330   &           &  1.83    &  5.2      &  11.1    &  0.84  &  0.68   &               \\
070411   &  2.954    &  27.00   &  13.9     &  49.7    &  0.77  &              &  15$^{a}$      \\
070419A  &  0.97     &  5.58    &  2.1      &  1.0     &  1.04  &              &  2.45$^{a}$    \\
070420   &           &  140.00  &  48.0     &  235.9   &  0.72  &  0.67   &               \\
070506   &  2.31     &  2.10    &  8.1      &  18.2    &  0.83  &              &  3.0$^{a}$     \\
070518   &           &  1.62    &  3.3      &  4.0     &  0.91  &  0.15   &               \\
070529   &  2.4996   &  25.70   &  8.2      &  15.1    &  0.86  &              &  15.8$^{a}$    \\
070611   &  2.04     &  3.91    &  11.4     &  8.5     &  0.98  &              &  1.5$^{a}$     \\
070616   &           &  192.00  &  49.0     &  20.5    &  1.06  &  0.85   &               \\
070628   &           &  35      &  139.4    &  430.0   &  0.79  &  0.92   &               \\
070721A  &           &  0.71    &  0.7      &  4.0     &  0.70  &  1.1    &               \\
070810A  &  2.17     &  6.9     &  9.1      &  4.8     &  1.03  &              &  5.1$^{a}$     \\
070911   &           &  120     &  32.5     &  174.0   &  0.71  &  0.73   &               \\
070917   &           &  20      &  6.6      &  47.9    &  0.67  &  14     &               \\
071003   &  1.60435  &  83      &  180.1    &  123.7   &  0.99  &  0.48   &  2.2$^{f}$     \\
071010A  &  0.98     &  2.0     &  43.0     &  51.6    &  0.92  &  4      &  16.79$^{\otimes}$   \\
071010B  &  0.947    &  44      &  48.3     &  324.6   &  0.68  &              &  2.4$^{a}$     \\
071011   &           &  22      &  26.6     &  76.4    &  0.80  &  3.4    &               \\
071020   &  2.145    &  23      &  13.1     &  106.6   &  0.65  &  0.4    &  4.44$^{\otimes}$    \\
071031   &  2.692    &  9.0     &  16.6     &  2.1     &  1.22  &              &  1.4$^{a}$     \\
071112C  &  0.823    &  30      &  5.6      &  17.3    &  0.79  &              &  1.2$^{a}$     \\
071122   &  1.14     &  5.8     &  3.7      &  1.2     &  1.09  &              &  1.8$^{a}$   \\
\\
UL& & & & & & &\\
------------& & & & & & &\\
\\
$\beta_{\rm OX} <$ 0.5 & & & & & & &\\
------& & & & & & &\\
\\
050713B$^{\tablenotemark{\circledast}}$  &           &  31.80   &  2.26     &  387.6   &  0.24  &  2.15   &               \\
050716$^{\tablenotemark{\circledast}}$   &           &  61.70   &  0.32     &  24.4    &  0.35  &  0.8    &               \\
060306   &           &  21.30   &  1.2      &  59.2    &  0.41  &  3.69   &               \\
060319$^{\tablenotemark{\circledast}}$   &           &  2.64    &  0.20     &  38.7    &  0.23  &  4.06   &               \\
060719   &           &  15.00   &  0.1      &  16.9    &  0.24  &  3.91   &               \\
061202   &           &  34.20   &  7.7      &  270.9   &  0.46  &  5.12   &               \\
070219   &           &  3.19    &  0.1      &  7.6     &  0.35  &  3.19   &               \\
070223$^{\tablenotemark{\circledast}}$   &           &  17.00   &  0.35     &  20.0    &  0.39  &  4.5    &               \\
070412   &           &  4.84    &  0.1      &  4.3     &  0.43  &  1.24   &               \\
070521   &  0.553    &  80.10   &  0.2      &  106.0   &  0.09  &  5.53   &  13.94$^{\otimes}$   \\
\\
0.5 $\leq \beta_{\rm OX} <$ 0.6 & & & & & & &\\
------& & & & & & &\\
\\
050128$^{\tablenotemark{\circledast}}$   &           &  50.20   &  7.87     &  158.2   &  0.53  &  0.091  &               \\
050502B\tablenotemark{\ast}  &           &  4.78    &  2.16     &  42.6    &  0.53  &  0.082  &               \\
050803   &  0.422    &  21.50   &  8.2      &  124.7   &  0.57  &  1.34   &  26.5$^{b}$    \\
050922B  &           &  22.30   &  5.3      &  69.5    &  0.59  &  0.91   &               \\
060814   &  0.84     &  146.00  &  8.9      &  175.5   &  0.53  &  2.89   &  10.40$^{\otimes}$   \\
060904A  &           &  77.20   &  3.0      &  68.2    &  0.52  &  1.43   &               \\
060923C  &           &  15.80   &  0.8      &  10.5    &  0.59  &  0.79   &               \\
061004   &           &  5.66    &  0.4      &  6.0     &  0.57  &  0.12   &               \\
070103   &           &  3.38    &  0.3      &  4.0     &  0.59  &  2.71   &               \\
070328   &           &  90.60   &  9.8      &  188.6   &  0.54  &  2.02   &               \\
070517   &           &  2.15    &  1.0      &  13.1    &  0.59  &  0.39   &               \\
070621   &           &  43      &  1.0      &  19.8    &  0.53  &  4.4    &               \\
\\
$\beta_{\rm OX} \geq$ 0.6 & & & & & & &\\
------& & & & & & &\\
\\
050117   &           &  88.10   &  207.1    &  409.0   &  0.85  &              &               \\
050124$^{\tablenotemark{\circledast}}$   &           &  11.90   &  89.3     &  51.3    &  1.02  &              &               \\
050219A  &           &  41.10   &  11.7     &  24.3    &  0.84  &  0.48   &               \\
050223   &  0.5915   &  6.36    &  7.7      &  4.5     &  1.01  &              &               \\
050326   &           &  88.60   &  20.3     &  97.5    &  0.73  &  1.44   &               \\
050410   &           &  41.50   &  14.3     &  10.6    &  0.98  &  0.01   &               \\
050412$^{\tablenotemark{\circledast}}$   &           &  6.18    &  0.19     &  1.5     &  0.66  &  0.17   &               \\
050421   &           &  1.45    &  23.3     &  1.9     &  1.28  &  4.6    &               \\
050422   &           &  6.07    &  1955.2   &  3.0     &  1.82  &  0.94   &               \\
050509A  &           &  3.41    &  2.7      &  9.7     &  0.77  &  2.3    &               \\
050528$^{\tablenotemark{\circledast}}$   &           &  4.48    &  3.79     &  8.8     &  0.83  &              &               \\
050714B  &           &  5.95    &  3.3      &  6.3     &  0.85  &  3.27   &               \\
050717   &           &  63.10   &  14.3     &  3.3     &  1.14  &  1.6    &               \\
050726   &           &  19.40   &  9.7      &  18.5    &  0.85  &  0.095  &               \\
050819   &           &  3.50    &  4.2      &  6.1     &  0.89  &  0.18   &               \\
050822   &           &  24.60   &  8.6      &  68.3    &  0.66  &  0.65   &               \\
050827   &           &  21.00   &  6.7      &  72.9    &  0.62  &  2.23   &               \\
050916   &           &  9.29    &  16739.7  &  23.6    &  1.83  &  1.2    &               \\
051001   &           &  17.40   &  2.1      &  7.2     &  0.77  &  1.53   &               \\
051016A  &           &  8.37    &  9.7      &  1.5     &  1.19  &  3.42   &               \\
051021B  &           &  8.35    &  813.1    &  5.3     &  1.63  &  0.48   &               \\
051117B  &           &  1.77    &  12.4     &  1.6     &  1.22  &  2.4    &               \\
060105   &           &  176.00  &  44.1     &  47.6    &  0.93  &  1.85   &               \\
060109   &           &  6.55    &  35.4     &  18.4    &  1.03  &  2.28   &               \\
060123   &  1.099    &  3.0     &  99.7     &  63.0    &  1.00  &  0.99   &               \\
060211A  &           &  15.70   &  5.5      &  10.5    &  0.85  &  0.45   &               \\
060211B  &           &  4.38    &  2.8      &  5.2     &  0.86  &  0.55   &               \\
060219   &           &  4.28    &  0.9      &  1.9     &  0.84  &  3.51   &               \\
060312   &           &  19.70   &  652.4    &  16.3    &  1.44  &  1.02   &               \\
060322   &           &  52.20   &  68.6     &  13.5    &  1.16  &  1.9    &               \\
060403   &           &  13.50   &  223.0    &  9.6     &  1.37  &  2.5    &               \\
060413   &           &  35.60   &  6702.5   &  266.5   &  1.38  &  6.9    &               \\
060421   &           &  12.50   &  6173.4   &  12.5    &  1.78  &  2.3    &               \\
060427   &           &  4.99    &  1.3      &  4.3     &  0.78  &  0.92   &               \\
060501   &           &  12.20   &  4.8      &  2.7     &  1.02  &  14     &               \\
060515   &           &  14.10   &  24.0     &  3.2     &  1.21  &  2340   &               \\
060602A$^{\tablenotemark{\circledast}}$  &  0.787    &  51.23   &  48.0     &  21.0    &  1.06  &              &               \\
060712   &           &  12.40   &  17.5     &  9.3     &  1.03  &  1.46   &               \\
060717   &           &  0.65    &  1.1      &  1.6     &  0.89  &  2.6    &               \\
060805A  &           &  0.72    &  4.8      &  2.1     &  1.05  &  0.65   &               \\
060807\tablenotemark{\star }   &           &  4.94    &  16.84    &  37.3    &  0.83  &  1.05   &               \\
060813   &           &  54.60   &  98.9     &  169.8   &  0.87  &  0.64   &               \\
060825   &           &  4.53    &  1.09     &  15.9    &  0.58  &  0.63   &               \\
060919   &           &  5.46    &  29.8     &  3.3     &  1.24  &  9.3    &               \\
060923B  &           &  4.80    &  91.2     &  23.0    &  1.13  &  3      &               \\
060929$^{\tablenotemark{\circledast}}$   &           &  8.30    &  1.58     &  3.8     &  0.82  &  1.05   &               \\
061002   &           &  6.77    &  11.0     &  2.5     &  1.14  &  2      &               \\
061028   &           &  9.66    &  7.2      &  2.7     &  1.07  &  0.34   &               \\
061102   &           &  2.79    &  11.0     &  0.2     &  1.49  &  0.86   &               \\
070107   &           &  51.70   &  42.4     &  147.9   &  0.77  &  0.5    &               \\
070220$^{\tablenotemark{\circledast}}$   &           &  14.71   &  1.60     &  33.8    &  0.83  &  1.53   &               \\
070227   &           &  16.20   &  49.7     &  22.5    &  1.05  &  1      &               \\
070429A  &           &  9.10    &  17.6     &  28.0    &  0.88  &  0.38   &               \\
070509   &           &  1.75    &  3.5      &  2.6     &  0.98  &  4.2    &               \\
070520A  &           &  2.50    &  1.5      &  0.8     &  1.03  &  2.35   &               \\
070520B  &           &  9.25    &  4.0      &  2.1     &  1.03  &  1.28   &               \\
070531$^{\tablenotemark{\circledast}}$   &           &  10.80   &  5.0      &  3.4     &  0.99  &  0.22   &               \\
070612B  &           &  16.80   &  47.0     &  21.9    &  1.04  &  4.7    &               \\
070704   &           &  126.5   &  1.952    &  22.4    &  1.18  &  4      &               \\
070714A  &           &  1.5     &  3.2      &  8.5     &  0.81  &  4.7    &               \\
070805   &           &  7.2     &  25.5     &  8.7     &  1.09  &              &               \\
070808$^{\tablenotemark{\circledast}}$   &           &  12      &  1.88     &  3.0     &  0.88  &  8.3    &               \\
070920B  &           &  6.6     &  6.0      &  9.5     &  0.88  &              &               \\
071028A  &           &  3.0     &  24.6     &  8.8     &  1.08  &  0.18   &               \\
071101   &           &  0.76    &  415.3    &  1.6     &  1.70  &  2.6    &               \\
 \enddata
\tablenotetext{\S}{$R$-band and X-ray fluxes mainly adopted from Nysewander et al. (2008)
unless specified otherwise.}
\tablenotetext{\dag}{http://www.mpe.mpg.de/$^\sim$jcg/grbgen.html;
http://swift.gsfc.nasa.gov/docs/swift/archive/grb$\_$table.html/.}
\tablenotetext{\ddag}{http://swift.gsfc.nasa.gov/docs/swift/archive/grb$\_$table.html/;
Sakamoto et al. 2008} \tablenotetext{\ast}{050502B; supplemented by us:
http://www.swift.ac.uk/xrt$\_$curves/00116116/; Prabhu (2005).}
\tablenotetext{\star}{060807; supplemented by us:
http://www.swift.ac.uk/xrt$\_$curves/00223217/; Duscha et al. (2006).}
\tablenotetext{\circledast}{Deeper optical observations of Melandri et al. (2008)
were adopted to replace the Nysewander et al. (2008) values.}
\tablenotetext{\oplus}{References for $N_{\rm H}$:
[a] http://www.swift.ac.uk/xrt$\_$spectra/ (all $N_{\rm H}$ @ $z=0$ values unless specified otherwise);
[b] Grupe et al. 2007; [c] Goad et al. 2005; [d] Troja et al. 2007; [e] Jaunsen et al. 2008; [f] Perley et al. 2008.}
\tablenotetext{\otimes}{Values derived from $N_{\rm H}$ @ $z=0$ multiplied by $(1+z)^{2.1}$.}
\end{deluxetable}

\begin{deluxetable}{lccccccccc}
 \tabcolsep 0.5mm
 %\tabletypesize{\scriptsize}
 %\tablecolumns{9}
 \tablewidth{0pt}
 %\tablenum{1}
 \tablecaption{$Swift$ high-redshift GRB catalog\label{tab:HighZEvents}}
 \tablehead{
 \colhead{ID}  &  \colhead{Redshift} &  \colhead{Fluence\tablenotemark{\ddag}}  &  \colhead{Flux Density} &  \colhead{Flux Density\tablenotemark{\dag}} &  \colhead{$\beta_{\rm OX}1$}  &  \colhead{Flux Density\tablenotemark{\dag\dag}} &  \colhead{$\beta_{\rm OX}2$}  & \colhead{Intrinsic $N_{\rm H}$\tablenotemark{\oplus}} & \colhead{ \tablenotemark{\S} }\\
 \colhead{}  &  \colhead{$z$}  &  \colhead{[15-150 keV]} &  \colhead{3keV @ 11h} &  \colhead{$R$ @ 11h} &  \colhead{@ 11h }  &  \colhead{$R_{\rm ext}$ @ 11h} &  \colhead{@ 11h }  & \colhead{@ GRB $z$}  & \colhead{Ref}\\
 \colhead{}  &  \colhead{}  &  \colhead{10$^{-7}$ erg cm$^{-2}$} &  \colhead{10$^{-3} \mu$Jy} &  \colhead{$\mu$Jy} &  \colhead{}  &  \colhead{$\mu$Jy} &  \colhead{}  & \colhead{10$^{21}$ cm$^{-2}$}  & \colhead{}
 }
 \startdata
%#ID       & z     & Flu   & FDX    & FDR   &  B1    &  FDR   & B2   & NH_Int & Ref
050505    & 4.27  & 24.90 & 117.7 & 4.7   &  0.50  &  8.0   & 0.57   & 24.0$^{b}$   & 1,2 \\
050730    & 3.967 & 23.80 & 273.6 & 15.2  &  0.55  &  33.7  & 0.66   & 12.4$^{b}$   & 1,3 \\
050814    & 5.3   & 20.10 & 42.4 & 2.1   &  0.53  &  16.2  & 0.81    & 7.3$^{a}$    & 1,4 \\
050904    & 6.295 & 48.30 & 1.3 & $<$2.48  &  $<$1.03 &  20.0  & 1.31   & 39.3$^{b}$   & 1,5 \\
060206    & 4.048 & 8.31  & 63.2 & 82.7  &  0.98  &  101.9 & 1.01    & 5.1$^{a}$    & 1,6 \\
060210    & 3.91  & 76.60 & 267.9 & 1.3   &  0.21  &  2.88  & 0.29   & 19.5$^{b}$   & 1,7 \\
060223A   & 4.41  & 6.73  & 1.2 & 3.6   &  1.09  &    & $>$1.09     & 27.0$^{a}$   & 1 \\
060510B   & 4.9   & 40.70 & 4.3 & 0.4   &  0.62  &  1.32  & 0.78     & 45.5$^{b}$   & 1,8 \\
060522    & 5.11  & 11.40 & 8.9 & 3.7   &  0.82  &    & $>$0.82     & 21.0$^{a}$   & 1 \\
060927    & 5.47  & 11.30 & 3.7 & 0.315 &  0.60  &  4.1   & 0.95     & 12.0$^{a}$   & 1,9 \\
%#080913A  & 6.7   & 5.6   & 0.0013 & 0.08  &  <0.56 &  1.0  & 0.90  & 34$^{g}$     & 1,10 \\
 \enddata
\tablenotetext{\ddag}{http://swift.gsfc.nasa.gov/docs/swift/archive/grb$\_$table.html/;
Sakamoto et al. 2008.}
\tablenotetext{\dag}{True $R$-band observational results.}
\tablenotetext{\dag\dag}{Values extrapolated from the NIR/IR observations.}
\tablenotetext{\oplus}{References for $N_{\rm H}$ are the same as in Table 1.}
\tablenotetext{\S}{References: [1] Nysewander M. et al. 2008; [2] Hurkett et al. 2006;
[3] Pandey et al. 2006; [4] Jakobsson et al. 2006; [5] Rumyantsev et al. 2005;
[6] Curran et  al. 2007b; [7] Curran et al. 2007a; [8] Melandri et al. 2006;
[9] Ruiz-Velasco et al. 2007.}
\end{deluxetable}

\label{lastpage}


\begin{thebibliography}{99}
\small \setlength{\itemindent}{-3mm} \setlength{\itemsep}{-0.5mm}
\setlength{\baselineskip}{4.5mm}


%% you can type \apj for ApJ, \aap for A&A, \apss for Ap&SS, etc. Please consult
%% the macro raa.cls. You can also find them in aasguide.tex (AASTeX for ApJ, AJ, PASP)
%% Please follow the formats of RAA's references list as demonstrated below:

%\bibitem{} Andersen, M. I., et al. 2000, A\&A, 364, L54

%\bibitem{} Barthelmy, S. D., et al. 1995, Ap\&SS, 231, 235

\bibitem{} Bal\'{a}zs, L. G., et al. 2008, in AIP Conf. Proc., 1000, 44 (arXiv:0903.5275)

\bibitem{} Basa, S., et al. 2008, in Proceedings of the Annual Meeting of
    the French Society of Astronomy and Astrophysics, eds. C. Charbonnel,
    F. Combes, \& R.Samadi, 161 (arXiv:0811.1154)

\bibitem{} Berger, E., et al. 2002, ApJ, 581, 981

%\bibitem{} Bessell, M. S. 2005, ARA\&A, 43, 293

%\bibitem{} Bloom, J., et al. 1999, MNRAS, 305, 763

\bibitem{} Butler, N. R., Kocevski, D. 2007, ApJ, 668, 400

%\bibitem{} Burrows, D. N., et al. 2005, Space Sci Rev, 120, 165

%\bibitem{} Castro-Tirado, A. J., et al. 2007, A\&A, 475, 101

\bibitem{} Cenko, S. B., et al. 2009, ApJ, 693, 1484

\bibitem{} Campana, S., et al. 2006, Nature, 442, 1008

%\bibitem{} Costa, E., et al. 1997, Nature, 387, 783

\bibitem{} Curran, P. A., et al. 2007a, A\&A, 467, 1049

\bibitem{} Curran, P. A., et al. 2007b, MNRAS, 381, L65

\bibitem{} Dai, Z. G., Cheng, K. S. 2001, ApJ, 558, L109

\bibitem{} Dai, Z. G. 2004, ApJ, 606, 1000

\bibitem{} De Pasquale, M., et al. 2003, ApJ, 592, 1018

%\bibitem{} De Pasquale, M., et al. 2008, ArXiv e-prints (arXiv:0809.4688)

\bibitem{} Dickey, J. M., Lockman, F. J. 1990, ARA\&A, 28, 215

\bibitem{} Djorgovski, S. G., et al. 2001, ApJ, 562, 654

\bibitem{} Duscha, S., et al. 2006, GCN Circ., 5417

%\bibitem{} Eichler, D., et al. 1989, Nature, 340, 126

\bibitem{} El\'{i}asd\'{o}ttir, \'{A}., et al. 2009, ApJ, 697, 1725

\bibitem{} Evans, P., et al. 2007, A\&A, 469, 379

%\bibitem{} Evans, P., et al. 2008, GCN Circ., 7955

\bibitem{} Fan, Y. Z., Piran, T. 2006, MNRAS, 369, 197

%\bibitem{} Frail, D. A., et al. 1997, Nature, 389, 261

%\bibitem{} Fruchter, A. S. 1999, ApJ, 512, L1

\bibitem{} Fruchter, A., Krolik, J. H., Rhoads, J. E. 2001, ApJ, 563, 597

\bibitem{} Fynbo, J., et al. 2001, A\&A, 373, 796

\bibitem{} Fynbo, J., et al. 2006, Nature, 444, 1047

%\bibitem{} Fynbo, J., et al. 2007, astro-ph/0703458

\bibitem{} Galama, T. J., Wijers, R. 2001, ApJ, 549, L209

\bibitem{} Gehrels, N., et al. 2004, ApJ, 611, 1005

\bibitem{} Gehrels, N., Cannizzo, J. K., Norris, J. P. 2007, NJPh, 9, 37

\bibitem{} Gehrels, N., et al. 2008, ApJ, 689, 1161

\bibitem{} Genet, F., et al. 2007, MNRAS, 381, 732

\bibitem{} Ghisellini, G., et al. 2007, ApJ, 658, L75

\bibitem{} Goad, M., et al. 2005, GCN Circ., 3952;

%\bibitem{} Greiner, J., et al. 2008, ArXiv e-prints (arXiv:0810.2314)

\bibitem{} Grindlay, J., et al. 2009, in AIP Conf. Proc., 1133, 18 (arXiv:0904.2210)

\bibitem{} Groot, P. J., et al. 1998, ApJ, 493, L27

\bibitem{} Grupe, D., et al. 2007, AJ, 133, 2216

\bibitem{} Hurkett, C. P., et al. 2006, MNRAS, 368, 1101

\bibitem{} Jakobsson, P., et al. 2004, ApJ, 617, L21

\bibitem{} Jakobsson, P., et al. 2005, ApJ, 629, 45

\bibitem{} Jakobsson, P., et al. 2006, A\&A, 447, 897

\bibitem{} Jaunsen, A. O., et al. 2008, ApJ, 681, 453

\bibitem{} Kann, D. A., et al. 2006, ApJ, 641, 993

%\bibitem{} Kann, D. A., et al. 2007, ArXiv e-prints (arXiv:0712.2186)

\bibitem{} Kann, D. A., et al. 2008, ArXiv e-prints (arXiv:0804.1959)

\bibitem{} Klose, S., et al. 2003, ApJ, 592, 1025

\bibitem{} Kr\"{u}hler, T., et al. 2008, ApJ, 685, 376

\bibitem{} Lamb, D. Q., Reichart, D. E. 2000, ApJ, 536, 1

\bibitem{} Lamb, D. Q., et al. 2004, NewAR, 48, 423

\bibitem{} Lazzati, D., et al. 2002, MNRAS, 330, 583

\bibitem{} Li, A., et al. 2008, ApJ, 685, 1046

\bibitem{} Liang, E. W., et al. 2007, ApJ, 670, 565

\bibitem{} Liang, E. W., et al. 2008, ApJ, 675, 528

\bibitem{} Lin, Y. Q. 2006, \chjaa, 6, 555

%\bibitem{} MacFadyen, A., Woosley S., 1999, ApJ, 524, 262

\bibitem{} Melandri, A., et al. 2006, GCN Circ., 5103;

\bibitem{} Melandri, A., et al. 2008, ApJ, 686, 1209

%\bibitem{} M\'{e}sz\'{a}ros, P., Ress, M. J., 1997, ApJ, 476, 232

%\bibitem{} Narayan, R., et al. 1992, ApJ, 395, L83

\bibitem{} Nardini, M., et al. 2008, ApJ, 386, L87

\bibitem{} Nysewander, M., et al. 2009, ApJ, 701, 824

\bibitem{} Oates, S. R., et al. 2006, MNRAS, 372, 327

%\bibitem{} Paczy\'{n}ski, B. 1986, ApJ, 308, L43

\bibitem{} Paczy\'{n}ski, B. 1998, ApJ, 494, L45

\bibitem{} Panaitescu, A. 2008, ArXiv e-prints (arXiv:0811.1235)

\bibitem{} Pandey, S. B., et al. 2006, A\&A, 460, 415

\bibitem{} Pedersen, K., et al. 2006, ApJ, 636, 381

\bibitem{} Perley, D. A., et al. 2008, ApJ, 688, 470

\bibitem{} Perley, D. A., et al. 2009, ArXiv e-prints (arXiv:0905:0001)

\bibitem{} Pian, E., et al. 2006, Nature, 442, 1011

%\bibitem{} Piran, T. 1999, Phys. Rep., 314, 575

%\bibitem{} Piro, L., et al. 2002, ApJ, 577, 680

\bibitem{} Prabhu, T. O. 2005, GCN Circ., 3346

\bibitem{} Rol, E., et al. 2005, ApJ, 624, 868

\bibitem{} Rol, E., et al. 2007, ApJ, 669, 1098

%\bibitem{} Roming, P. W., et al. 2005, Space Sci Rev, 120, 95

\bibitem{} Roming, P. W., et al. 2006, ApJ, 652, 1416

\bibitem{} Roming, P. W., et al. 2008, AAS, HEAD meeting, 10, 28.22

\bibitem{} Ruiz-Velasco, A., E., et al. 2007, ApJ, 669, 1

\bibitem{} Rumyantsev, V., et al. 2005, GCN Circ., 3939

%\bibitem{} Sakamoto, T., et al. 2006, GCN Circ., 5886

\bibitem{} Sakamoto, T., et al. 2008, ApJS, 175, 179

\bibitem{} Sari, R., et al. 1998, ApJ, 497, L17

\bibitem{} Schady, P., et al. 2007, MNRAS, 377, 273

\bibitem{} Schlegel, D., et al. 1998, ApJ, 500, 525

\bibitem{} Shao, L., Dai, Z. G. 2007, ApJ, 660, 1319

%\bibitem{} Spearman, C. 1904, Am J Psychol, 15, 72

\bibitem{} Starling, R., et al. 2007, ApJ, 661, 787

\bibitem{} Stratta, G., et al. 2004, ApJ, 608, 846

%\bibitem{} Stratta, G., et al. 2007, ApJ, 661, L9

\bibitem{} Tanvir, N. R., et al. 2008, MNRAS, 388, 1743

%\bibitem{} Taylor, G. B., et al. 1998, ApJ, 502, L118

\bibitem{} Troja, E., et al. 2007, ApJ, 665, 599

\bibitem{} Urata, Y., et al. 2007, PASJ, 59, L29

%\bibitem{} van Paradijs, J., et al. 1997, Nature, 386, 686

\bibitem{} van der Horst, A. J., et al. 2009, ApJ, 699, 1087

\bibitem{} Wang, X. Y., et al. 2007, ApJ, 664, 1026

\bibitem{} Waxman, E., Draine, B. T. 2000, ApJ, 537, 796

\bibitem{} Watson, D., et al. 2006, ApJ, 652, 1011

%\bibitem{} Woosley, S. 1993, ApJ, 405, 273

\bibitem{} Woosley, S., Bloom J. 2006, ARA\&A, 44, 507

%\bibitem{} Yost, S. A., et al. 2007, ApJ, 669, 1107

\bibitem{} Yu, Y. W., Dai, Z. G. 2007, A\&A, 470, 119

\bibitem{} Zhang, B. 2007a, \chjaa, 7, 1

%\bibitem{} Zhang, B., Fan, Y. Z., et al. 2006, ApJ, 642, 354

\bibitem{} Zhang, B., M\'{e}sz\'{a}ros, P. 2004, IJMPA, 19, 2385

\bibitem{} Zhang, B., et al. 2007b, ApJ, 655, 989

\bibitem{} Zheng, W. K., et al. 2008, \chjaa, 8, 693

\end{thebibliography}
\end{document}